# How happy are my neighbours? Modelling spatial spillover effects of well-being

**Authors:** Thanasis Ziogas[1,2], Dimitris Ballas[1], Sierdjan Koster[1], Arjen Edzes[1]

[1]Department of Economic Geography, Faculty of Spatial Sciences, University of Groningen, the Netherlands

[2]Corresponding author: a.ziogas@rug.nl

**Abstract**

This article uses data of subjective Life Satisfaction aggregated to the community level in Canada and examines the spatial interdependencies and spatial spillovers of community happiness. A theoretical model of utility is presented. Using spatial econometric techniques, we find that the utility of community, proxied by subjective measures of life satisfaction, is affected both by the utility of neighbouring communities as well as by the latter's average household income and unemployment rate. Shared cultural traits and institutions may justify such spillovers. The results are robust to the different binary contiguity spatial weights matrices used and to the various econometric models. Clusters of both high-high and low-low in Life Satisfaction communities are also found based on Moran's I test.

**Keywords:** spatial econometrics; spatial inter-dependencies; well-being spillovers

# Introduction

Over the past four decades [1] there has been significant progress in the measurement and analysis of the social, economic and demographic determinants of subjective measures of happiness and well-being. It has now long and convincingly been argued that subjective measures of life satisfaction and happiness are reliable [2,3,4]. There has also been a considerable academic and policy-related interest in the analysis of what affects happiness and well-being, ranging from individual demographic (e.g. age, gender, household type) and socio-economic factors (e.g. income, education level) to social, environmental and spatial context (e.g. quality of the environment, natural and human-made amenities) [4,5,6,7,8].

At the same time there has been a rapidly increasing and influential body of theoretical and empirical work highlighting the impact of inequality upon happiness and well-being [6, 9,10,11,12,13]. Yet, there is a relative paucity of research that explores the impact of interpersonal and inter-community social relations and of the impact of social and spatial inequalities. Notable exceptions include studies that considered social norms and attempted to model and quantify the effect of interactions between variables at different levels, such as the relationship between being unemployed and regional unemployment rate [14,15,16] or to use average regional income as a proxy to infer neighbourhood effects [17]. There have also been efforts to empirically consider social comparison effects by identifying and modelling reference groups [1,18].

There have been even more limited efforts to comprehensively consider social and spatial issues and possible spatial interdependencies and spillovers. This may be due to the significant data and methodological challenges regarding this type of socio-spatial analysis. In particular, there are very limited sources of data on subjective life satisfaction at the local level. In most cases the smallest area at which data is typically available and analysed is the region [19] or local authority district [16,20].

However, recent research published in this journal [21] addressed some of these challenges by combining suitable social survey and health survey microdata with small area data to build a new public use dataset for community-level life satisfaction in Canada, they identifying important inter-community differences. This new data set also opens up new possibilities for further comprehensive analysis that can provide insights regarding the importance of space, place and space for happiness and life satisfaction. As the researchers who created this data set point out (also inviting more researchers to use their novel new data set), possible 'next steps' and challenges include the examination of plausible sources of the substantial inter-community differences revealed in their original article as well as the analysis of social context variables. In this article we take up the challenge of examining these sources by focusing on possible spatial inter-dependencies and spillovers. In particular, we present a theoretical and methodological framework aimed at addressing the following questions:

*Does the collective sense of well-being of an area affect that of a neighbouring area?*
*Does the level of unemployment and other socio-economic indicators in one area affect that of neighbouring communities?*

## Background

There is a large body of academic literature exploring the role of relative social status and inter-personal social comparisons pertaining to a wide range of disciplines. This includes theoretical work on conspicuous consumption highlighting the importance of relative social status [22] as well as the theory of "bandwagon effect" in which mob motivation and mass psychology are taken into account in demand theory [23]. Also of relevance is the "relative income hypothesis" proposing that individual consumption functions depend not only on an individual's own consumption but also on the consumption of other individuals [24]. Other influential work highlighted the importance of social comparisons and reference groups [25] and relative deprivation [26].

More recently, there has also been very impactful work by social epidemiologists with compelling empirical evidence regarding 'status anxiety' [27] and of the detrimental impact of inequality upon social and individual well-being, resulting from psycho-social processes and social-evaluative threats [12,13].

The importance of social comparisons has also been extensively considered by the literature on the economics of happiness. Of particular relevance is the Easterlin paradox: many countries experienced high levels of growth in their GDPs but at the same time the levels of happiness for these countries remained constant [28,29,30]. The impact of inequality and social comparisons has been acknowledged as an important determinant of happiness and as a possible explanation for this paradox [1,9,10,31,32].

Nevertheless, there has been a relatively limited number of empirical studies examining the impact of social comparisons on happiness. Most of these studies involved an analysis of comparisons of individual or household income or wealth or consumption to that of the rest of a country in which an individual resides in [1,18,33,34]. There have also been efforts to perform similar analysis at the sub-national level with the inclusion of regional or sub-regional average income as an explanatory variable in happiness regression models as a proxy for social comparisons [17,35,36,37]. Another approach involves the examination of interactions between individual level and regional or sub-regional level characteristics [14,15,16] finding that "unemployment hurts, but it hurts less when there are more unemployed people around" [14, p.346]. More recent work also considered the impact of wellbeing inequality at the regional level upon individual well-being finding that higher inequality is associated with lower levels of well-being [38,39]. There have also been studies that used multilevel modelling approaches to implicitly capture the impact of social context upon (including social comparisons) upon happiness [16].

Nevertheless, all of these studies considering social comparisons and happiness tend to only focus on individual geographical units (whether these are countries, regions or sub-regional units or community areas) without attempting to analyse possible spatial interactions, spillovers or interdependencies between geographical units. To the best of our knowledge, there is only one peer-reviewed study to date that explicitly involved analysis of spatial relationships and spillover between small areas and this was possible due the availability of high quality microdata sets made available via special license for the north of the Netherlands [40]. The paucity of studies considering geography, space, place, socio-economic characteristics and happiness is to some extent due to the lack of publicly available data for geographical units smaller than the region. As also noted in the introduction, a significant positive development towards changing this has been a study published in this journal last year which we briefly discussed in the introduction. This study made available a new public use dataset for community-level life satisfaction in Canada, inviting other researchers to use it for further analysis. In this article we utilize the power of this new data set to make a contribution to the literature on social (and socio-spatial) comparisons and happiness by adopting a spatial econometrics approach.

## Data and method

# Theoretical considerations

To assess the role of socio-spatial comparisons in happiness levels, we enhance a standard utility model to include spatial spillover effects. The concept of utility has long been established in Economics in order to describe consumer behaviour and it has now long been suggested that happiness can be considered a suitable proxy to it [1,18,41]. Individual utility functions describe the elements that give utility to individuals. However, individuals consume many different goods that add to their utility. The concept of goods is expanded to include individual attributes and circumstances such as being in employment, in good health, owning a house etc. Given that there is an infinite number of goods that an individual can consume, we assume for simplicity that an individual's utility depends only on some personal characteristics. The variables often used to account for variation in utility include income, employment status, marital status, gender and age among others. Building on the idea of comparative life satisfaction which was discussed above, there is a need to include socio-spatial interdependencies in a standard utility model, building on previous pertinent theoretical work [33,42,43,44,45].

When we empirically estimate the utility function, the latter is of the form:

$$U_i = f(x_i, y_i, z_i \dots), \quad (1)$$

where U is a happiness index usually proxied by subjective measures of life satisfaction, while the elements x, y and z represent the income, employment status, marital status etc. For convenience, we denote all individual characteristics of individual *i* as **X**$_i$.

Previous theoretical and empirical work included the development of models that expand the textbook specification of utility and allow for externalities [14,46] enabling the explicit modelling of the ways in which the actions of one agent affect the utility of another agent. Under this scenario, the new utility function that accrues includes also characteristics of others that are likely to affect one's utility. Equation (1) is now of the form:

$$U_i = f(X_i, X_j), \quad (2)$$

where $X_j$ represents the characteristics of other individuals that affect individual *i*. It should also be noted that $i \neq j$.

Equation (2) can be further expanded in order to allow for some new form of externalities. Individuals interact with each other and based on this interaction they affect and they are affected. The influence that one individual has upon others is not limited to her own characteristics. It could be argued that an individual's utility might also help explain the utility of others. Hence, not only the income or the employment status of others may have a say upon her utility but also their utility per se, might play some role. Having said that, we need to modify equation (2) to account for this kind of externality. Hence, the new form is:

$$U_i = f(X_i, X_j, U_j), \quad (3)$$

where $U_j$ represents the utility of others and again $i \neq j$.

An important research challenge is to identify a way to empirically measure the impact of $U_j$ on $U_i$. In the following section, where we discuss the methodological framework, we provide the spatial econometric specification under which we can estimate this coefficient.

The main difference between the empirical part of the current article and the theory is that the latter is mostly focused on individual utility, whereas in our dataset we have aggregate (community level) measures of community's life satisfaction. Since the welfare of a community consists of the welfare of its individuals, we thus generalize the theory discussed above to describe the utility of communities. Therefore, we use the subscript *i* to denote a community instead of an individual. Accordingly the subscript *j* that used to denote the *others*, now denotes the *other communities*. At the community level however, the variables that might exert an influence towards the life satisfaction of another region should be modified accordingly. Hence, instead of having the personal income, employment status, age etc. as in the case of individual level, we now use the average income of the households, the unemployment rate of the community, and some geographical characteristics like whether the latter is rural or urban along with other community level variables. Having a dataset with the aforementioned variables and the geographical structure required for our purposes, we are able to empirically measure the interdependencies found in community level utilities. A crucial issue that needs to be considered is how 'others' are defined and specified from a modelling perspective. In theory, both at the individual and community area level, with *others* we include all those units that may influence one's utility. The next section details how we empirically establish the relevant others.

## Data

As also previously noted, we adopt a cross-sectional dataset for community life satisfaction in Canada which was made available as part of a study published in this journal [21]. The data set was built on by combining various waves from the Canadian Community Health Survey (CCHS) and the General Social Survey (GSS) in order to construct Life Satisfaction estimates for each of 1,215 Canadian communities based on more than 500,000 individuals in total. This is a high-resolution dataset on life satisfaction with the coordinates of the communities being attached. Based on this approach, Canada is divided into 1,215 geographic regions, both urban and rural. These 1,215 communities are a result of natural, built and administrative boundaries combined with a minimum sampling threshold of 250 individuals in each community to minimize the idiosyncratic component at the individual level [21]. Both the threshold of 250 individuals in the new synthesized populations and the assigned boundaries were selected by taking into account the Modifiable Areal Unit Problem (MAUP). The surface area of the communities is unevenly distributed as is evident from Fig 1. This reflects the uneven geographical dimensions in Canada with sparsely populated communities in the North and centre of the country and more densely populated areas on the coasts and around the cities. For example, the North-West territories (comprising an area of 1,173,793 km$^2$) are divided in three communities in our sample whereas Guelph covers an area of 593.51 km$^2$ and is divided into 5 distinct communities in the dataset. The different sizes of the communities likely affect the interdependencies between them. For this reason, we apply two different versions of the model in which the second focuses on smaller scale units.

**Fig 1. Communities in Canada.**

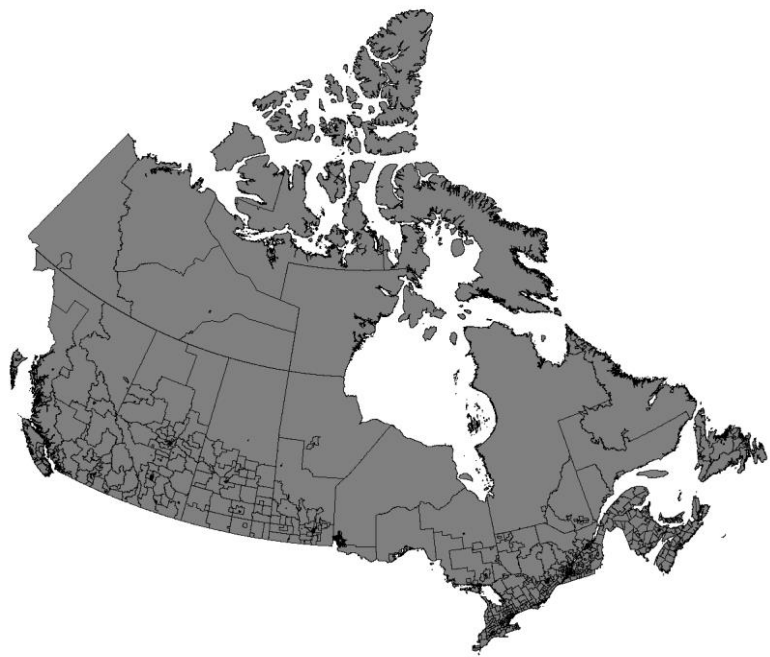

In the dataset for community life satisfaction in Canada, the measure of a communities' life satisfaction is based on aggregated individual measurements. As such, changes in life satisfaction can be attributed to changes in the community context as well as changes in the composition of the population following migration. In our empirical setup this is not expected to influence results as migration flows are relatively modest [47]. Also, we adopt a cross-sectional approach which is not sensitive to migration over time. It is also relevant to note here that 62.4% of the population in our sample lives for at least 5 years in the same house.

Regarding the Life Satisfaction variable per se, in both surveys, the question that individuals had to answer in order to record their well-being was of the form "Taking all things considered, how satisfied you are currently with your life on a scale ranging from 0 to 10" with 10 being the highest level of satisfaction. There is an on-going debate on whether such measures provide adequate consistency. Even though some economists might raise concerns about the statistical reliability of such measures, there have been many high impact evidence-based studies that confirm their validity [2,3,48]. For example, there are studies that show how such measures correlate with the blood pressure [49] and electroencephalogram measures [50]. Even within those that are in favour of subjective measures of life satisfaction, there is no consensus in applied research on whether such measures are truly ordered or cardinal in nature and which usage is appropriate [51,52]. However, in the absence of a better alternative and acknowledging the shortcomings of such measures, we argue that they can provide powerful insights into the debate about "What makes people better-off?" and particularly "Does the interaction between units (e.g. individuals, firms, communities, countries etc) spur any spatial spillovers in subjective well-being?" The latter question is the underpinning of interdependencies both between individuals and communities. It should be noted that the likely mechanisms underpinning possible spatial interactions and interdependencies are very much dependent on the scale of spatial units. Our research focuses on the spatial level of community as defined above. Apart from life satisfaction in each community, the dataset contains various community area variables. Table 1 summarizes the variables used in the empirical analysis. The variables represent common confounders in explaining community life satisfaction including both socio-economic characteristics of the population (income, unemployment, educational level) and contextual information (commuting, density). Importantly, the data allow for including the variance in life-satisfaction in the form of standard deviation. Life Satisfaction is the only dependent variable used in the regression models while the rest of the variables serve as independent variables.

**Table 1. Description of the variables.**

| Name of the variable | Description of the variable |
| --- | --- |
| Life Satisfaction: | Mean life satisfaction (survey weighted). |
| Household Income (log): | Mean household income (log transformed). |
| Unemployment Rate: | Unemployment rate, percent (%). |
| Commute Duration: | Median commute time measured in minutes. |
| Population Density (log): | Population density (log transformed). |
| Proportion of Religious: | Proportion of respondents affiliating with a religion. |
| Permanent Location (5y): | Proportion of households who have resided in the same place for 5 years or more. |
| Proportion of 4y degree: | Proportion of population aged 25-64 with a post-secondary certificate or degree. |
| Proportion Foreign Born: | Proportion of population not born Canadian citizens. |
| Std Dev. of Life Satisfaction: | Standard deviation of Life Satisfaction variable. |
| Urban: | Dummy variable equal to 1 for regions characterized as urban. |

Of particular interest is the standard deviation of Life Satisfaction, given the clear evidence (especially in highly developed economies) of a negative relationship between average well-being and well-being inequality [38,39]. We use the standard deviation to capture the inequality of life satisfaction within communities. A high standard deviation (variance) of happiness in a community, means that there are both high and low levels in life satisfaction individuals in this community. We argue that the inequality of life satisfaction in a community exerts an (negative) effect upon the general level of happiness and can be considered a community area characteristic. For that reason, we consider it as a potential variable that explains the life satisfaction of neighbouring communities as well.

## Spatial Weights Matrix

The approach we adopt originates from a regional economics perspective. The conceptualization of neighbours is based on a spatial weights matrix denoted by W. A weights matrix describes the spatial arrangement of units in a sample [53,54. It is always a squared N×N ($W_{N \times N}$) matrix that reflects the spatial connectivity among spatial units. Specifically, the dimension of the matrix we are using is 1215x1215 as the number of the communities. Among the weights matrices that are most often used in empirical research are the p-order binary contiguity matrix and the inverse distance matrix. In the former case, assuming that p=1, if two communities share a common border they are assigned as neighbours while all other communities are not neighbours. In matrix terms, this will lead to the value of 1 in the cell that represents these two communities, while the value of 0 will be assigned to non-neighbours. On the other hand, the inverse distance matrix has as its elements the inverse of the distance between each pair of communities. It is based on what has been described as the first law of geography which states that the mutual influence of spatial units dissipates with distance [55]. Irrespective of the method used to construct the matrix, each element in the main diagonal of the matrix is always zero as it represents the connectivity of a community with itself. For all the empirical analysis, the weights matrices have been normalized and no cut-off point has been used for the inverse distance matrix. Specifically, we have row normalized the binary contiguity weights matrices so as the sum of each row to equal one, while we have divided all elements of the inverse distance matrix with the largest eigenvalue of the matrix [56]. We are using both matrices and variations of them when we explore the data in order to examine the sensitivity of our results of the results to the formulation of the weight-matrix. We argue that a binary contiguity matrix can model better the underlying mechanisms that lead to the interdependencies in well-being compared to the inverse distance. It is reasonable to assume that a region can affect another region next to it rather than one at the other side of the country. Figs 2 and 3 give an example of how the different weights matrices are visualized in the map of Canada for a particular region (dark grey).

**Fig 2. Visualization of a 1$^{st}$ order binary contiguity spatial weights matrix.**

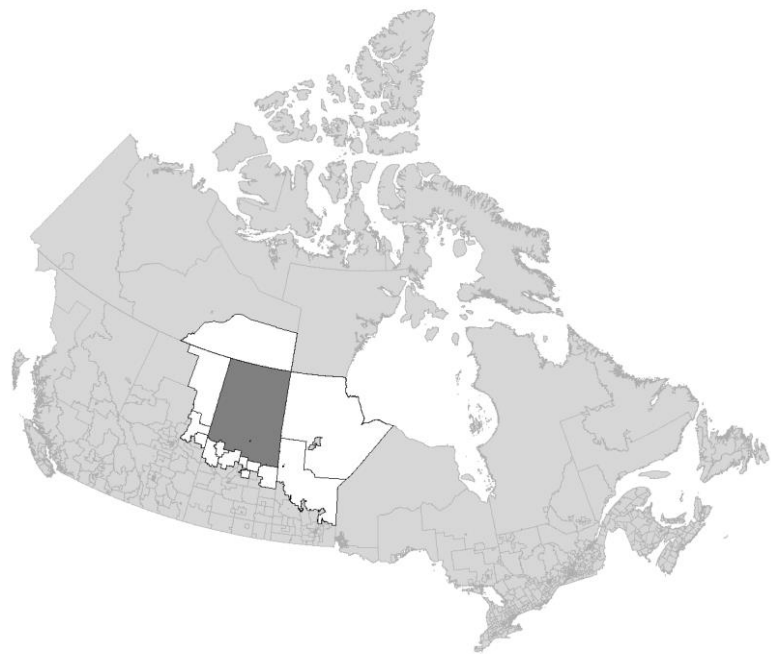

**Fig 3. Visualization of an inverse distance spatial weights matrix.**

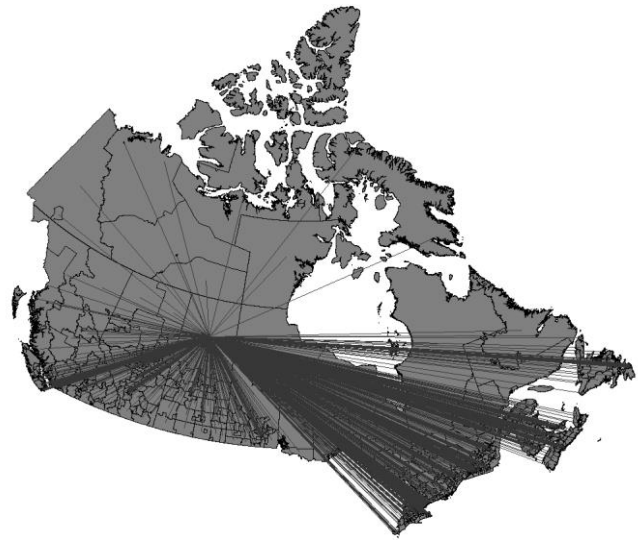

## Spatial Econometrics

Spatial econometrics is an expansion of the standard Ordinary Least Squares (OLS) regression in order to include spatial interactions using the weights matrix W. The general specification is as follows:

$$Y_i = \rho W Y_i + X_i B + W X_i \theta + \alpha_i + e_i,$$

(4)

$$e_i = \lambda W e_i + u_i$$

Equation (4) contains all possible spatial interactions: with the dependent variables (ρWYi), independent variables (WXθi) and with the error term (λWei) and is known as the General Nesting Spatial model (GNS). Different combinations of spatial interactions give rise to different spatial econometric models. Table 2, gives a brief description of the models.

**Table 2. Different combinations of spatial econometric models.**

| Name of model | Spatial interactions | Type of spatial spillovers |
|---|---|---|
| SAR, Spatial AutoRegressive model | WY | Constant ratios |
| SEM, Spatial Error Model | Wu | Zero by construction |
| SAC, Spatial Autoregressive Combined model | WY, Wu | Constant ratios |
| SLX, Spatial Lag of X model | WX | Fully flexible |
| SDEM, Spatial Durbin Error Model | WX, Wu | Fully flexible |
| SDM, Spatial Durbin Model | WX, WY | Fully flexible |
| GNS, General Nesting Spatial model | WX, WY, Wu | Fully flexible |

Following [56], only models that include a spatial lag of X (WXθi) are able to produce flexible spatial spillover effects. Spatial spillovers are of interest in spatial econometrics as they show how a change in an explanatory variable of region *j*, impacts the dependent variable in unit *i* (≠*j*). The definition of a spatial spillover is the marginal impact of a change to one independent variable in a one cross-sectional unit on the dependent variable in another unit. This is known as the indirect effect. Direct effects on the other hand, measure the impact of a change of an explanatory variable of spatial unit *i* on the dependent variable of the same unit *i*. These estimations are derived from the reduced form of the spatial econometric model at hand. The estimation is usually carried out using Maximum Likelihood. Equation (5) presents the reduced form for the SDM model:

$$Y_i = (I - \rho W)^{-1}[X_i B + W X_i \theta + \alpha_i + e_i] \quad (5)$$

Given that only models that include spatial interactions with the independent variables are able to produce flexible spatial spillovers effects we focus our analysis in such models. Apart from that, since our intention is to estimate the interdependencies in utilities, we are particularly interested in the coefficient of $U_j$ from Equation (3) which is the same as the coefficient rho (ρ) in Equation (4). For that reason, we prefer models that include spatial interactions with the dependent variable as well. The significance threshold we follow is that of 5% (denoted in the tables by two asterisks), however, we also consider both the 10% and 1% significance levels (denoted with one and three asterisks respectively).

We also apply Moran's I test in order to estimate both the sign of the spatial autocorrelation and to detect whether there are clusters of high-high, low-low or mixed in life satisfaction communities. In addition, the same test is used along with different spatial weights matrices to examine whether adopting spatial econometric techniques is indeed the appropriate next stage of analysis. The software programs used for the current article are GeoDa 1.14.0, ArcMap 10.5.1 and Stata 16 [57,58,59].

# Results

## Descriptive analysis

The main variable of interest is the Life Satisfaction of the regions. It has already been stated that we use this variable to proxy utility which in term represents the welfare of the communities. Hence, we start by presenting some descriptive statistics that provide insights for the regression analysis employed later in the article. Fig 4 presents the distribution of Life Satisfaction and Fig 5 the standard deviation of Life Satisfaction, both figures along with the normal and kernel function distributions. The mean of Life Satisfaction in our sample is 8.04 on a 0 to 10 scale. This is in line with the results in the OECD better life index, which ranks Canada as one of the countries with the most satisfied people [60]. Both variables behave as normally distributed variables since both normal and kernel functions are close together. In Fig 4, however, we observe a negative skewness (-0.35) while positive skewness (0.31) is detected in Fig 5 implying that there are some outliers, i.e.

communities with very low life satisfaction compared to the average and communities with high life satisfaction inequality compared to the average.

**Fig 4. Distribution of Life Satisfaction.**

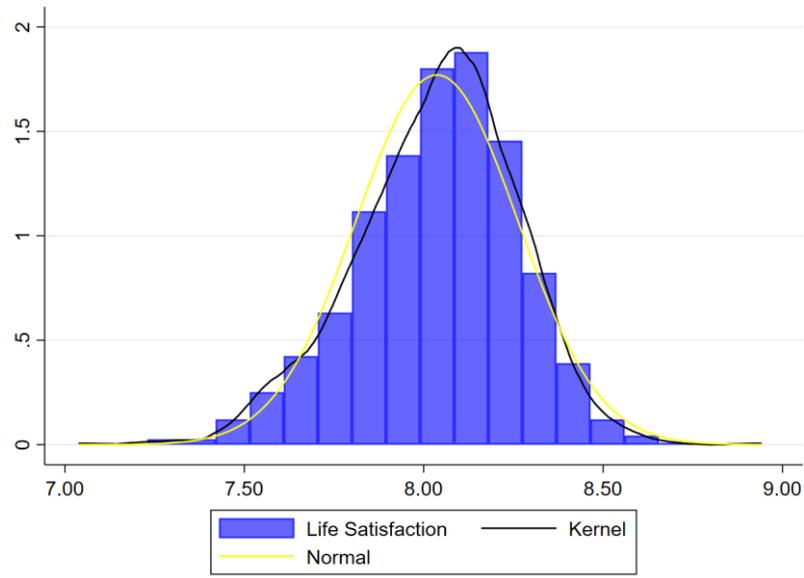

**Fig 5. Distribution of community-level inequality (std. dev.) in Life Satisfaction.**

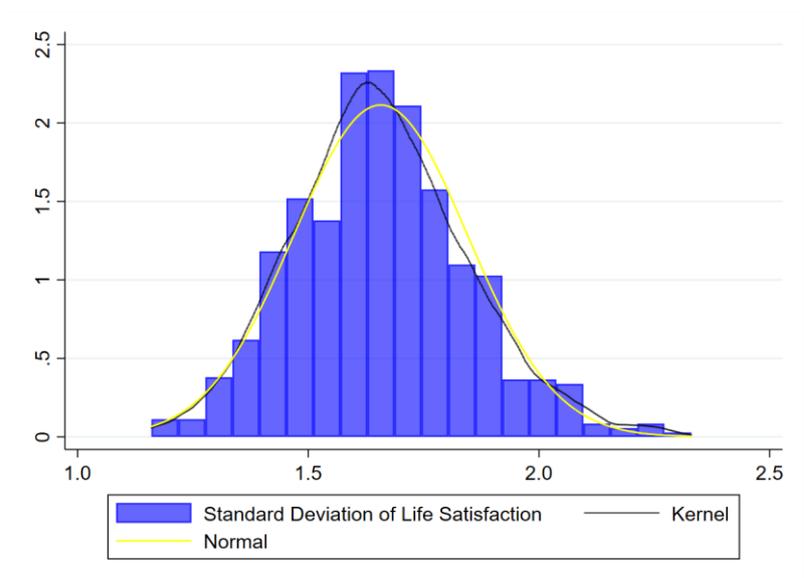

We proceed by making a first exploration towards the geographical aspects of the communities under examination. It has often been suggested in the literature that rural areas are more satisfied with life than urban ones [61] and especially in more affluent countries [4,39]. Our results are also consistent with this. Fig 6, shows the difference in life satisfaction between urban and rural areas and Fig 7 the differences in standard deviation of Life Satisfaction. There is a statistically significant difference (t-statistics equals 14) in Life Satisfaction between the two areas: urban areas record a value of 7.97 and rural of 8.15. On the other hand, the difference in the dispersion of Life Satisfaction between the two kinds of regions is less striking and not significant at 5% significance level (t-statistics equals 1,753).

**Fig 6. Distribution of Life Satisfaction for Urban and Rural areas.**

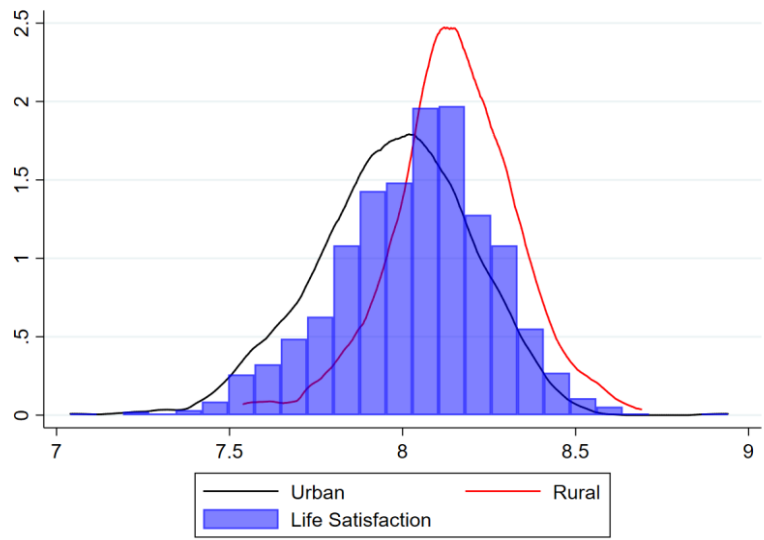

**Fig 7. Distribution of community-level inequality (std. dev.) in Life Satisfaction for Urban and Rural areas.**

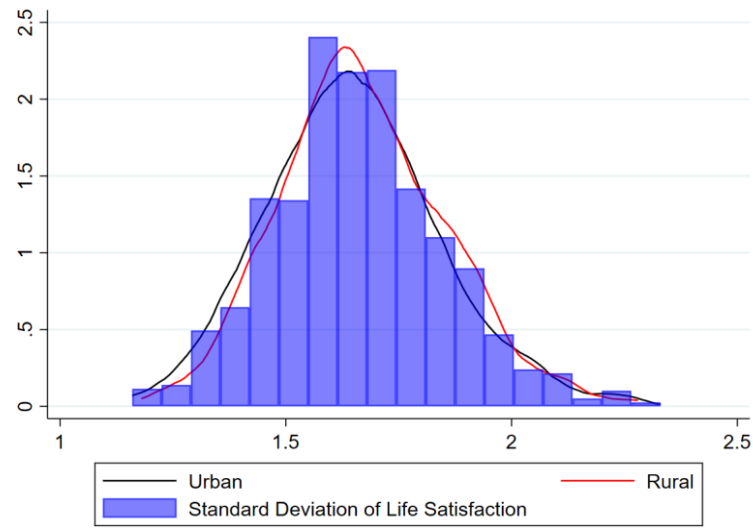

An important observation is the negative association found between life satisfaction and the standard deviation of life satisfaction. One would argue that the more spread out the life satisfaction in a region is, or the more spatial inequality in happiness a community contains, the less satisfied are the individuals that reside in it. Fig 8, presents this observation for the full sample while Fig 9 for the rural-urban sub-samples. The negative association is more intense in urban areas. Hence, it is not only that urban areas are in general less satisfied with life but also the dispersion of happiness within those areas is higher, suggesting that greater life satisfaction inequality can be found in urban places. This is consistent with relevant recent research presented in the World Happiness Report [39]. A possible explanation for this observation is the difference on how densely populated those two areas are. The mean value of the logarithm of population density in urban areas is 6.92 while for rural is 2.29. As a result, in urban areas inequality is more evident since more individuals reside in these areas and comparisons are more likely to take place.

**Fig 8. Scatter plot between Life Satisfaction and community-level inequality (std. dev.) in Life Satisfaction for the entire sample.**

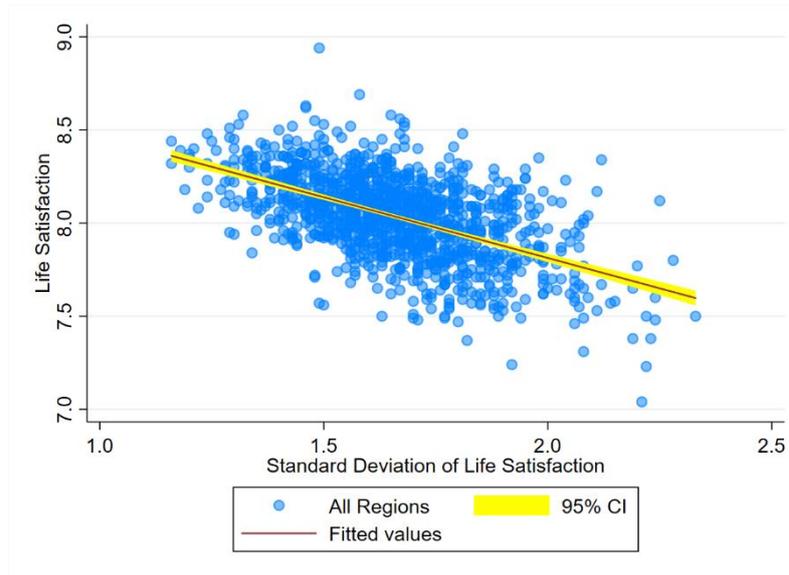

**Fig 9. Scatter plot between Life Satisfaction and community-level inequality (std. dev.)Life Satisfaction for the urban-rural subsamples.**

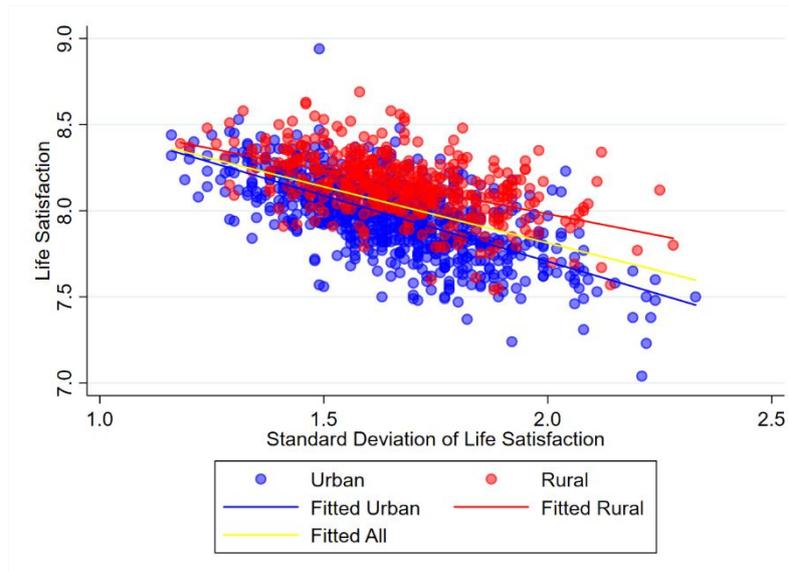

Fig 10 presents the spatial distribution of Life Satisfaction across Canadian communities. Life Satisfaction appears to be spatially dependent since the high satisfied communities appear to be clustered together. Particularly, it seems that there are many communities in the east of the country (Quebec) with high levels of life satisfaction. This observation suggests that cultural or institutional characteristics may be in place in these communities that explain this pattern. Spatial statistics can further explore the data and shed light to the underlying mechanisms governing these observations.

**Fig 10. Spatial Distribution of Life Satisfaction across 1215 Canadian communities.**

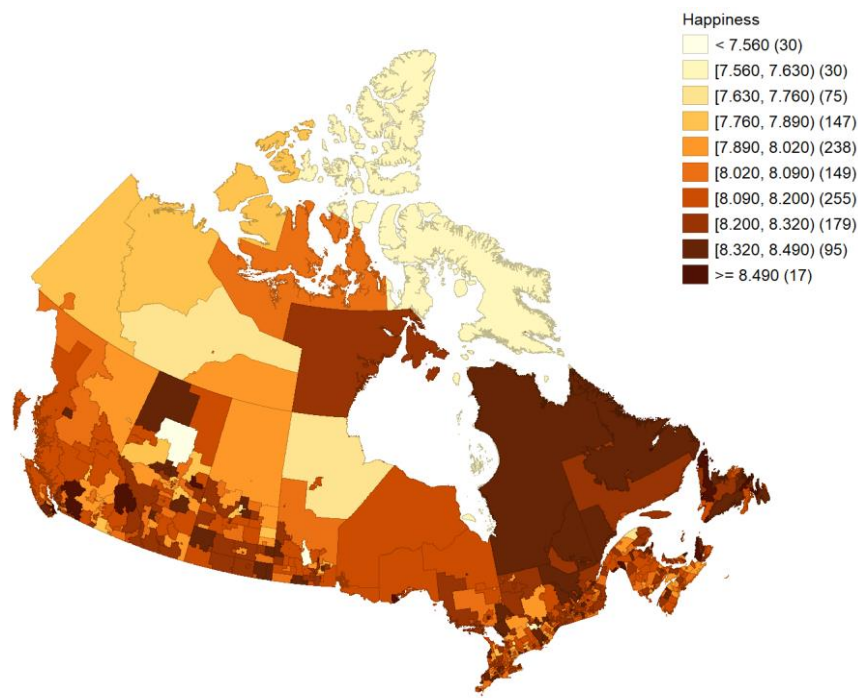

Fig 11 presents the spatial distribution of standard deviation of Life Satisfaction. Unlike the level of life satisfaction, the spatial clustering regarding standard deviation is less striking. Yet, the negative association between the two variables is also visible in the maps as the darker colours in Fig 10, that represent high life satisfaction communities, are now depicted with bright colours (lower variation). Once more, this is a visualization of the important finding that in more satisfied communities the variance of life satisfaction is lower compared to less satisfied communities where variance is high and hence more unequal.

**Fig 11. Spatial Distribution of inequality (std. dev.) In life Satisfaction across 1215 Canadian communities.**

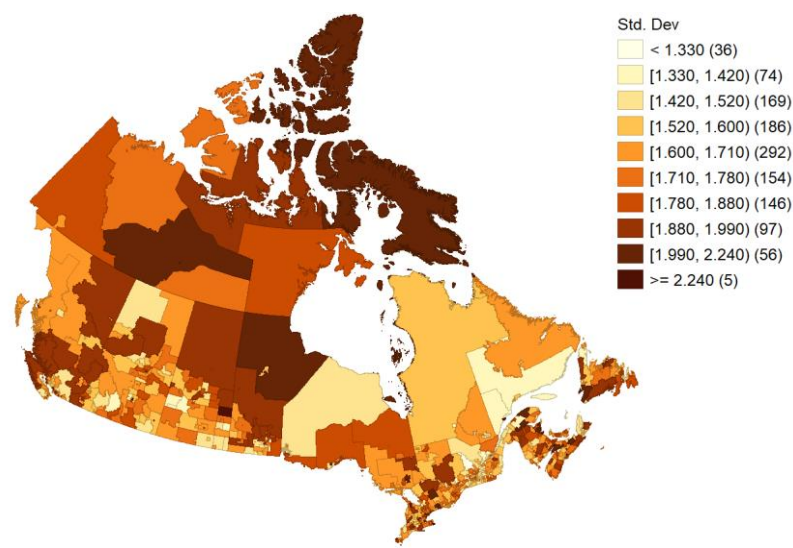

The information provided by the correlation matrix is valuable since we get a first intuition about the relationship between our variables in all possible combinations. A general comment is that the majority of the correlations in Table 3, are statistically significant. First of all, there is a positive and significant correlation at 1% level between Life Satisfaction and the average Household Income. In contrast, there is a negative correlation between Life Satisfaction and Unemployment Rate as well as with standard deviation of Life

Satisfaction. The former correlation is statistically significant at 10% level while the latter at 1% level. Apart from that, it is also interesting to note that there is a positive correlation between Urban areas, Commute Duration and Population Density as one might expect. The explanation for that is straightforward since the majority of economic activity is often concentrated in urban areas which leads to overpopulation resulting in traffic jams and increased commute duration. Finally, a thought-provoking correlation is observed between the community-level education attainment (Proportion of 4y degree) and community income (Household Income) on the one hand which is positive and significant and between the former and community level of Unemployment Rate (negative and significant) on the other.

**Table 3. Correlation Matrix.**

|  | (1) | (2) | (3) | (4) | (5) | (6) | (7) | (8) | (9) | (10) | (11) |
|---|---|---|---|---|---|---|---|---|---|---|---|
| (1) Life Satisfaction | 1.000 | | | | | | | | | | |
| (2) Household Income | 0.1051*** | 1.000 | | | | | | | | | |
| (3) Unemployment Rate | -0.0507* | -0.3932*** | 1.000 | | | | | | | | |
| (4) Commute Duration | -0.2654*** | 0.3501*** | -0.1730*** | 1.000 | | | | | | | |
| (5) Population Density | -0.4465*** | 0.1257*** | -0.1667*** | 0.4241*** | 1.000 | | | | | | |
| (6) Proportion of Religious | 0.3078*** | -0.1754*** | 0.1938*** | -0.0256 | -0.1577*** | 1.000 | | | | | |
| (7) Permanent Location | 0.4041*** | -0.0194 | 0.1857*** | -0.0579** | -0.5069*** | 0.4524*** | 1.000 | | | | |
| (8) Proportion of 4y degree | -0.0250 | 0.5174*** | -0.3395*** | 0.3580*** | 0.5245*** | -0.1672*** | -0.3238*** | 1.000 | | | |
| (9) Proportion Foreign Born | -0.5077*** | 0.2417*** | -0.0138 | 0.5706*** | 0.5980*** | -0.2419*** | -0.3636*** | 0.3634*** | 1.000 | | |
| (10) Std Dev. of Life Satisfaction | -0.5464*** | -0.2532*** | 0.2642*** | -0.0650** | -0.0068 | -0.0914*** | -0.0509* | -0.3293*** | 0.0945*** | 1.000 | |
| (11) Urban | -0.3730*** | 0.3084*** | -0.2099*** | 0.5025*** | 0.7729*** | -0.1599*** | -0.3854*** | 0.5317*** | 0.5018*** | -0.0503* | 1.000 |

* $p<0.1$, ** $p<0.05$, *** $p<0.01$

## Spatial Autocorrelation

There are several diagnostic tests available that formally examine whether the residuals from the OLS regression are spatially correlated. If this is the case, we need to employ spatial regression models. In order to be able to examine spatial correlation, we first need to estimate a simple OLS model using our standard explanatory variables. Using then the weights matrices we discussed in the Weights matrix section, we apply those diagnostics. Table 4, presents the OLS regression while Table 5 carries the test diagnostics on whether we should continue with error lag or spatial lag model

**Table 4. Life Satisfaction Regression, OLS, No weights matrix used.**

| **Dependent Variable**: Life Satisfaction | | | |
|---|---|---|---|
| **Independent Variables:** | Coefficient | t-statistic | $p$-value |
| Household Income (log) | 0.091 | 4.06 | 0.000 |
| Unemployment Rate | 0.002 | 1.25 | 0.211 |
| Commute Duration | -0.004 | -4.58 | 0.000 |
| Population Density (log) | -0.017 | -6.89 | 0.000 |
| Proportion of Religious | 0.236 | 5.82 | 0.000 |
| Permanent Location (5y or more) | 0.320 | 4.82 | 0.000 |
| Proportion of 4y degree or higher | 0.199 | 3.22 | 0.000 |
| Proportion Foreign Born | -0.340 | -8.82 | 0.000 |
| Std Deviation of Life Satisfaction | -0.556 | -21.50 | 0.000 |
| constant | 7.642 | 30.65 | 0.000 |
| Number of obs.: 1215 | | $F(9, 1205)$: | 196.48 |
| Adj. R-squared: 0.611 | | Prob > F: | 0.000 |

Note: Robust standard errors are used

**Table 5. Diagnostic tests for spatial dependence in OLS regression**

| Weights Matrices | Queen 1st B.C. | | Queen 2st B.C. | | Rook 1st B.C. | | Inverse Distance | |
|---|---|---|---|---|---|---|---|---|
| Test | Statistic | $p$-value | Statistic | $p$-value | Statistic | $p$-value | Statistic | $p$-value |
| Spatial Error | | | | | | | | |
| Moran's I | 5.771 | 0.000 | 5.957 | 0.000 | 5.970 | 0.000 | 4.011 | 0.000 |
| Lagrange multiplier | 29.215 | 0.000 | 29.219 | 0.000 | 29.675 | 0.000 | 2.704 | 0.100 |
| Robust Lagrange multiplier | 28.998 | 0.000 | 29.151 | 0.000 | 30.967 | 0.000 | 2.844 | 0.092 |
| Spatial Lag: | | | | | | | | |
| Lagrange multiplier | 0.225 | 0.635 | 0.075 | 0.784 | 2.649 | 0.104 | 12.836 | 0.000 |
| Robust Lagrange multiplier | 0.008 | 0.929 | 0.006 | 0.936 | 3.941 | 0.047 | 12.977 | 0.000 |

The tests in Table 5 report that we cannot accept the hypothesis that the residuals based on the OLS specification are independent and identically distributed. Specifically, the test considered the alternative hypothesis that residuals are correlated with nearby residuals as defined by the different weights matrices. Namely all matrices point to the adoption of a model with spatial interaction in the error term while only for the first order binary contiguity matrix of a rook form and for the inverse distance matrix there is evidence for a model having spatial interaction with the dependent variable. However, based on the theory that we are interested in, we are dictated to use models that can estimate the coefficient rho ($\rho$) from Equation (4). It has already been stated that coefficient rho is the coefficient of $U_j$, the utility of others. Subsequently, in the following section, we present the results based on spatial econometric models.

Furthermore, using Moran's I test for spatial autocorrelation, we generate maps showing how the communities in the sample are clustered and whether there is any spatial dependence among them. Fig 12, shows the graphs for the spatial autocorrelation found in Life Satisfaction while Fig 13, carries the maps with clusters of communities for the two main weights matrices adopted.

**Fig 12. Spatial autocorrelation of Life Satisfaction on 1215 Canadian communities.** Panel (A) shows the 1st order binary contiguity matrix (ROOK), while panel (B) the inverse distance matrix.

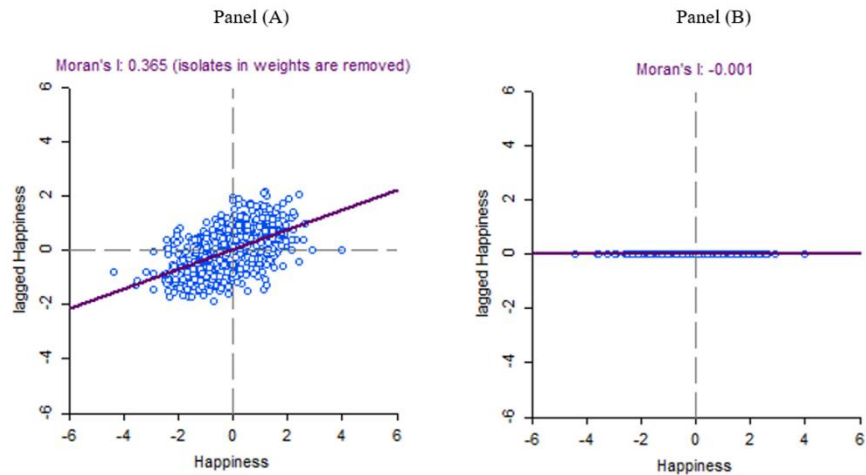

**Fig 13. Cluster and significance maps of spatial autocorrelation of Life Satisfaction on 1215 Canadian communities.** Panel (A) shows the 1st order binary contiguity matrix (ROOK), while panel (B) the inverse distance matrix.

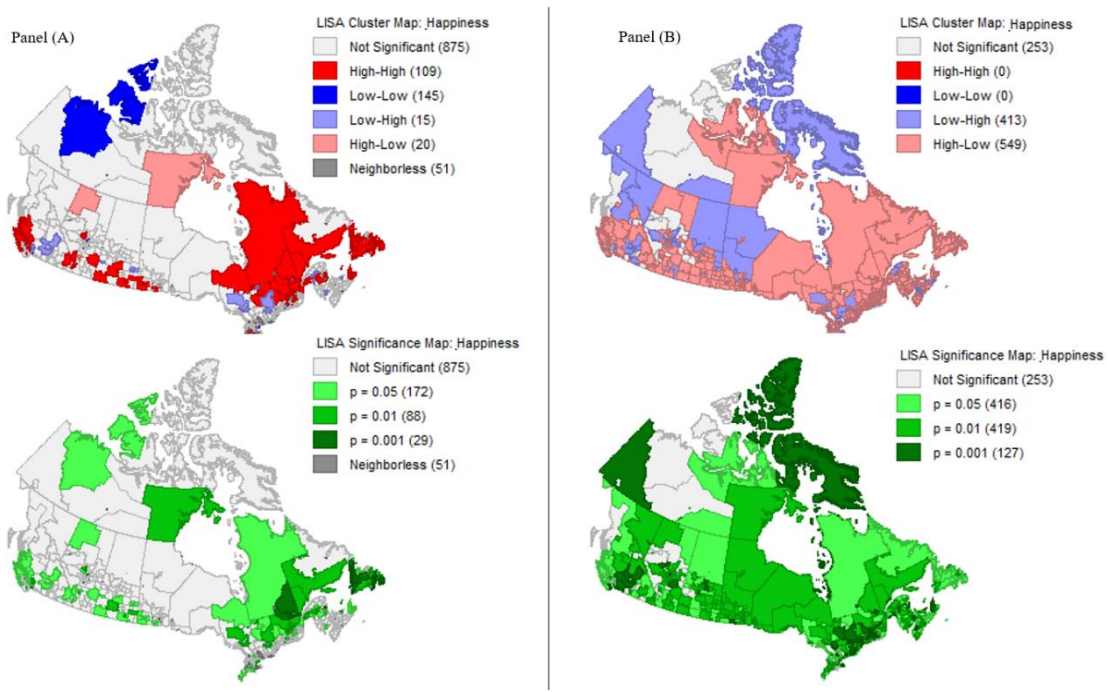

The results suggest that there is a positive spatial autocorrelation when we use the first order binary contiguity matrix in rook form, Fig 12, panel (A), and that there are many clusters of high-high in life satisfaction communities in Canada as the red colour in Fig 13, panel (A) shows. On the contrary, panels (B) in both figures indicate that there is no spatial autocorrelation in Life Satisfaction when we use the inverse distance matrix. The inverse distance matrix includes information from all the communities, although the ones further away carry less weight, and as a result possible spatial associations have arguably been averaged. Regardless, it shows that the spatial associations as fueled by interactions and social comparisons tend to have a relatively small spatial range.

## Spatial Regressions

We start our analysis using the SLX model for some preliminary exploration of the flexible spatial spillover effects. Subsequently, we employ models with more spatial interactions that are able to estimate the coefficient of $U^*$, for instance the Spatial Durbin Model (SDM) and the General Nested Model (GNS). We acknowledge the difficulty in obtaining the estimated coefficients from the latter model due to over-parameterization [62]. However, we use it for robustness checks. Tables 6 and 7, summarize the results. In the upper panel we present the direct effects, namely the impact that characteristics of a community have an influence upon its life satisfaction levels, while the middle panel presents the indirect effect. The latter is the average impact of the characteristics of neighbouring communities, as defined according to the weights matrix used, upon the life satisfaction of the community. Finally the bottom panel presents the two spatial regression coefficients. Because we are mostly interested in the indirect and spatial coefficients, for simplicity we include significance stars only in those coefficients. In both Tables 6 and 7, considering that it is a cross-sectional analysis, the R-square is large ranging from 0.611 to 0.618. All direct effects, regardless of the weights matrix and the model used, have the expected signs and are statistically significant. An exception from this rule is the coefficient of unemployment rate which is insignificant. There is a positive relationship between household income and life satisfaction, a result consistent in the literature that deals with individual level data either in developed or developing countries [63,64]. Both commuting time and population density are negatively associated with community's happiness as we saw in the correlation matrix before. This suggests, after correcting for other factors, that there are certain diseconomies to scale for the population. Finally, the proxy we used for the inequality of happiness, the standard deviation of life satisfaction, is negative and highly statistically significant (1% level). Again, it is clear that inequality hurts the satisfaction found in communities. For the indirect effects as well as for the two spatial coefficients, rho ($\rho$) and lambda ($\lambda$), some of the results differ based on the spatial weight matrix. In Table 6, we see that the life satisfaction level of a community is negatively affected by the income of the neighbouring communities, irrespective of the model, while the significance of the results ranges from 10% level to 1% level. Regarding the unemployment rate, we observe that there is a positive and significant indirect effect at the 1% level. The higher the unemployment rate in a community is, the more satisfied with life are the neighbouring communities. This finding may suggest some form of socio-spatial comparison effect, a possible sense of 'relief' or in other words an 'it could be worse' effect between communities, although more research is needed in this direction in order to establish a causal relationship. It is also worth highlighting that the coefficient rho has a positive and significant sign at 1% level in the SDM model and at 10% level at the GNS model. According to our theory, there is a positive association between neighbouring spatial units' utilities. Hence the first estimations we get for the coefficient of $\mathbf{U}_j$ are 0.162 and 0.087 for the SDM and GNS respectively.

**Table 6. Model Comparison of the estimated direct and spillover (marginal) effects on Life Satisfaction** (Weight Matrix Used: 1st Order B.C. Rook).

| | Dependent variable: Life Satisfaction | | | | |
|---|---|---|---|---|---|
| Independent (**Direct**) | OLS | SLX | SDEM | SDM | GNS |
| Household Income (log) | 0.091*** | 0.087*** | 0.093*** | 0.151*** | 0.121*** |
| | (4.06) | (4.33) | (4.17) | (6.53) | (4.50) |
| Unemployment Rate | 0.002 | -0.003* | -0.003* | -0.002 | -0.002 |
| | (1.25) | (-1.79) | (-1.95) | (-1.08) | (-1.53) |
| Commute Duration | -0.004*** | -0.003*** | -0.003*** | -0.003*** | -0.003*** |
| | (-4.58) | (-3.97) | (-3.79) | (-3.09) | (-3.30) |
| Population Density (log) | -0.017*** | -0.017*** | -0.017*** | -0.014*** | -0.016*** |
| | (-6.89) | (-7.78) | (-7.33) | (-6.38) | (-6.43) |
| Proportion of Religious | 0.236*** | 0.216*** | 0.235*** | 0.167*** | 0.198*** |
| | (5.82) | (5.75) | (5.40) | (4.35) | (4.53) |
| Permanent Location (5y) | 0.320*** | 0.305*** | 0.267*** | 0.268*** | 0.267*** |
| | (4.82) | (5.49) | (4.57) | (4.86) | (4.66) |
| Proportion of 4y degree | 0.199*** | 0.175*** | 0.170** | 0.119** | 0.143** |
| | (3.22) | (2.89) | (2.59) | (1.96) | (2.19) |
| Proportion Foreign Born | -0.340*** | -0.342*** | -0.333*** | -0.270*** | -0.297*** |
| | (-8.82) | (-8.97) | (-7.69) | (-6.77) | (-6.46) |
| Std Dev. of Life Satisfaction | -0.556*** | 0.560*** | -0.559*** | -0.552*** | -0.557*** |
| | (-21.5) | (-23.3) | (-23.5) | (23.2) | (-23.4) |
| Independent (**Indirect**) | | | | | |
| Household Income (log) | | -0.011* | -0.011* | -0.124*** | -0.070** |
| | | (-1.77) | (-1.72) | (-5.28) | (-2.05) |
| Unemployment Rate | | 0.007*** | 0.007*** | 0.005*** | 0.006*** |
| | | (3.99) | (3.98) | (2.63) | (3.13) |
| Std Dev. of Life Satisfaction | | 0.007 | 0.010 | -0.020 | -0.006 |
| | | (0.18) | (0.25) | (-0.47) | (-0.13) |
| ρ | | | | 0.162*** | 0.087* |
| | | | | (5.37) | (1.82) |
| λ | | | 0.212*** | | 0.126** |
| | | | (5.50) | | (2.04) |
| Adj. R-squared: | 0.611 | 0.617 | 0.617 | 0.617 | 0.618 |

Note: Maximum Likelihood estimation using robust standard errors. Marginal effects are presented. t-statistics in parentheses. The weights matrix has been row-normalized.

* p<0.1, ** p<0.05, *** p<0.01

**Table 7. Model Comparison of the estimated direct and spillover (marginal) effects on Life Satisfaction (Weight Matrix Used: Inverse Distance).**

| | Dependent variable: Life Satisfaction | | | | |
|---|---|---|---|---|---|
| Independent (**Direct**) | OLS | SLX | SDEM | SDM | GNS |
| Household Income (log) | 0.091*** | 0.084*** | 0.083*** | 0.089*** | 0.082*** |
| | (4.06) | (4.08) | (3.78) | (4.05) | (3.54) |
| Unemployment Rate | 0.002 | 0.001 | 0.001 | 0.001 | 0.001 |
| | (1.25) | (0.56) | (0.47) | (0.63) | (0.45) |
| Commute Duration | -0.004*** | -0.002*** | -0.002** | -0.003*** | -0.002** |
| | (-4.58) | (-2.97) | (-2.51) | (-2.98) | (-2.50) |
| Population Density (log) | -0.017*** | -0.012*** | -0.011*** | -0.012*** | -0.011*** |
| | (-6.89) | (-4.93) | (-4.39) | (-4.94) | (-4.25) |
| Proportion of Religious | 0.236*** | 0.246*** | 0.252*** | 0.241*** | 0.254*** |
| | (5.82) | (6.29) | (6.03) | (6.02) | (5.86) |
| Permanent Location (5y) | 0.320*** | 0.312*** | 0.310*** | 0.313*** | 0.310*** |
| | (4.82) | (5.58) | (5.43) | (5.59) | (5.42) |
| Proportion of 4y degree | 0.199*** | 0.212*** | 0.220*** | 0.207*** | 0.222*** |
| | (3.22) | (3.46) | (3.47) | (3.35) | (3.44) |
| Proportion Foreign Born | -0.340*** | -0.253*** | -0.237*** | -0.240*** | -0.239*** |
| | (-8.82) | (-5.80) | (-4.93) | (-5.07) | (-4.57) |
| Std Dev. of Life Satisfaction | -0.556*** | 0.557*** | -0.557*** | -0.557*** | -0.557*** |
| | (-21.5) | (-22.9) | (-23.0) | (22.9) | (-23.0) |
| Independent (**Indirect**) | | | | | |
| Household Income (log) | | -0.009 | -0.014 | -0.031 | -0.008 |
| | | (-0.63) | (-0.76) | (-0.82) | (-0.16) |
| Unemployment Rate | | 0.018*** | 0.019*** | 0.018*** | 0.019*** |
| | | (3.00) | (2.62) | (2.93) | (2.62) |
| Std Dev. of Life Satisfaction | | -0.056 | -0.041 | -0.039 | -0.043 |
| | | (-0.55) | (-0.32) | (-0.35) | (-0.33) |
| $\rho$ | | | | 0.051 | -0.016 |
| | | | | (0.65) | (-0.13) |
| $\lambda$ | | | 0.513*** | | 0.525*** |
| | | | (2.82) | | (2.67) |
| Adj. R-squared: | 0.611 | 0.618 | 0.617 | 0.617 | 0.617 |

Note: Maximum Likelihood estimation using robust standard errors. Marginal effects are presented. t-statistics in parentheses. The weights matrix has been normalized by dividing each element with the largest eigenvalue of the matrix.

* $p<0.1$, ** $p<0.05$, *** $p<0.01$

In Table 7, where we used the inverse distance matrix, the results are less conclusive. There is no significant indirect income effect while the rho coefficient is not significant either, meaning that there are no spillover effects of life satisfaction among the communities when the assumed spatial structure for possible interactions is increased to include a large part of Canada. Once again, only the unemployment rate has a significant indirect effect as in Table 6. It is interesting that in both Tables 6 and 7, the indirect effect of life satisfaction inequality is not significant suggesting that apart from the direct effect, the inequality found in one community does not relate to the life satisfaction levels of neighbouring ones.

**Robustness**

Given the large size of some of the communities in Canada, especially in the northern part of the country, we redo the analysis on the neighbourhoods of an urban area where spatial units are smaller and hence are assumed to be more connected. The reason for this distinction is the geographical surface of the northern regions per se. The problem that arises in those large communities is that their centres are quite distant from each other and hence the interaction between the regions is probably less intense. In other words, individuals that reside in those large areas may rarely cross the borders of their own communities resulting in less interaction with individuals from neighbouring ones. On the contrary, between the neighbourhoods of an urban area, the interaction between the communities is expected to be higher. The same effect likely applies to the social comparison arguments, even though media and other digital communication opportunities also allow for social comparisons across larger distances. Thus, in this exercise, we examine whether the results remain robust when we restrict the analysis to smaller geographical units where interactions are likely to happen more often between the spatial units and in which, consequently, social comparisons are facilitated.

In Fig 14, we present the maps of the communities included in the robustness check. We use 355 (274) communities, all belonging to the province of Ontario. The 355 regions constitute the entire province of Ontario while the 274 accrue after having removed the "island" spatial units and the remote areas. Another advantage of this exercise is the more homogenous sample of regions in terms of surface area. Canada has both English and French speaking communities and hence there may be differences that could be attributed to cultural or linguistic effects [65]. By focusing in the regions of Ontario, we eliminate the cultural differences that might be responsible for results we found before. Furthermore, given the different provinces in the entire sample, by focusing in just one province, we achieve a more homogenous sample in terms of institutions. The institutional differences that might have played a role before are now eliminated. Thus, two types of robustness take place in this exercise, the first concerns all the rural and urban communities of Ontario and the second the urban and rural communities around the city of Toronto.

**Fig 14. Urban and Rural communities within Ontario and around Toronto.**

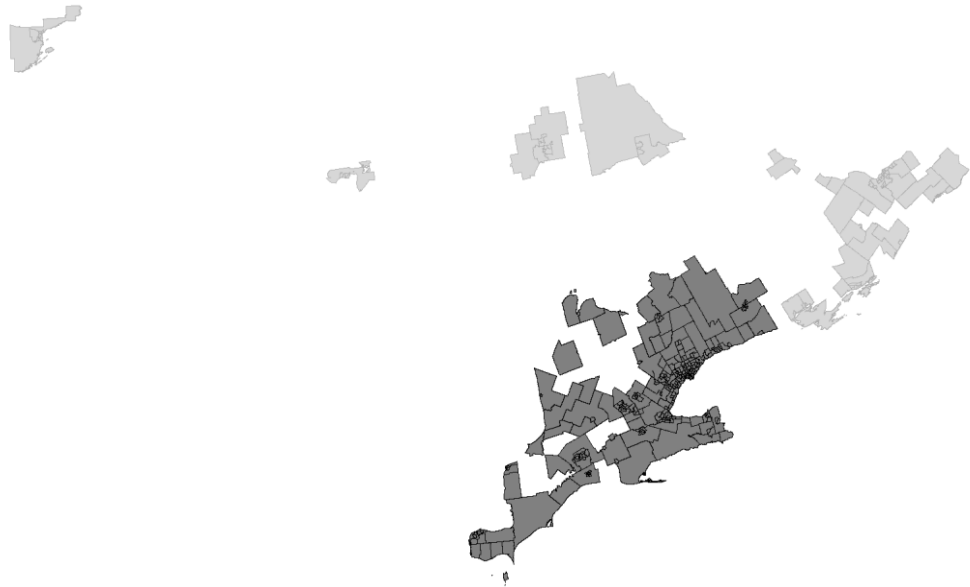

In Tables 8 and 9, we present the same regression as in Tables 6 and 7 but this time we are using the new restricted sample with the 274 regions while in Tables 10 and 11 we are using the 355 communities that constitute Ontario, namely those that include the "islands" and the remote areas. Once again, we examine all the different spatial econometric models and we use the two main spatial weights matrices, hence the binary contiguity matrix of the rook form and the inverse distance matrix for sensitivity analysis. In Tables 8-11 the direct effects are qualitatively the same as in all previous regressions. What is more important is the behaviour of the indirect effects. We observe some discrepancies compared to the full sample. While we were expecting to find stronger results in terms of magnitude and significance since communities within the same province were used, it turns out that the only variable that remains significant is the unemployment in most of the cases and especially in the model with the 255 regions. Particularly, unemployment exhibits a significant coefficient when we use the weights matrix of the rook form. In general, the results are less significant under the inverse distance matrix since under the latter conceptualization all regions are connected to some extent and hence the effect of neighbours evens out. This means that at the community level, the unemployment rate of one community is the most important predictor of the life satisfaction of neighbouring communities that share a common border. These results may have important policy implications and especially in relation to policies that affect unemployment because of the association between the latter and the happiness level of the population within neighbouring communities. Regarding the spatial coefficients, they are insignificant in most of the cases regardless of the model and the matrix used. R-square has been increased in both exercises compared to the entire sample, suggesting that the independent variables explain the variability in happiness in a better way. One major limitation of this exercise in the relatively small sample size that leads to higher standard errors of the coefficients compared to the entire sample and this may be the reason for the statistically insignificant results.

**Table 8. Model Comparison of the estimated direct and spillover (marginal) effects on Life Satisfaction. Restricted sample to Toronto communities (N=274) (Weight Matrix Used: 1st Order B.C. Rook).**

| | Dependent variable: Life Satisfaction | | | | |
|---|---|---|---|---|---|
| Independent (**Direct**) | OLS | SLX | SDEM | SDM | GNS |
| Household Income (log) | 0.137** | 0.172*** | 0.172*** | 0.193*** | 0.217*** |
| | (2.34) | (2.80) | (2.79) | (3.07) | (3.36) |
| Unemployment Rate | 0.001 | -0.006 | -0.006 | -0.006 | -0.004 |
| | (0.27) | (-0.98) | (-1.02) | (-0.94) | (-0.62) |
| Commute Duration | -0.007*** | -0.006*** | -0.006*** | -0.005*** | -0.004** |
| | (-3.39) | (-3.61) | (-3.54) | (-2.70) | (-2.07) |
| Population Density (log) | -0.036*** | -0.039*** | -0.038*** | -0.035*** | -0.033*** |
| | (-5.16) | (-5.63) | (-5.52) | (-4.92) | (-4.55) |
| Proportion of Religious | 0.382*** | 0.369*** | 0.374*** | 0.359*** | 0.328*** |
| | (3.15) | (2.98) | (2.98) | (2.89) | (2.81) |
| Permanent Location (5y) | 0.140 | 0.024 | 0.018 | -0.002 | -0.004 |
| | (1.04) | (0.18) | (0.14) | (-0.01) | (-0.03) |
| Proportion of 4y degree | 0.115 | 0.048 | 0.044 | 0.043 | 0052 |
| | (0.79) | (0.32) | (0.29) | (0.29) | (0.37) |
| Proportion Foreign Born | -0.224*** | -0.238*** | -0.241*** | -0.225*** | -0.201*** |
| | (-2.90) | (-2.85) | (-2.84) | (-2.68) | (-2.56) |
| Std Dev. of Life Satisfaction | -0.580*** | -0.574*** | -0.574*** | -0.576*** | -0.574*** |
| | (-11.97) | (-12.69) | (-12.69) | (-12.74) | (-12.75) |
| Independent (**Indirect**) | | | | | |
| Household Income (log) | | -0.020 | -0.019 | -0.078* | -0.140* |
| | | (-1.61) | (-1.56) | (-1.70) | (-1.80) |
| Unemployment Rate | | 0.016** | 0.017** | 0.014* | 0.010 |
| | | (2.11) | (2.16) | (1.77) | (1.21) |
| Std Dev. of Life Satisfaction | | 0.060 | 0.055 | 0.039 | 0.045 |
| | | (0.72) | (0.65) | (0.43) | (0.47) |
| P | | | | 0.096 | 0.186* |
| | | | | (1.40) | (1.72) |
| Λ | | | 0.036 | | -0.159 |
| | | | (0.38) | | (-1.06) |
| Adj. R-squared: | 0.686 | 0.695 | 0.695 | 0.697 | 0.698 |

Note: Maximum Likelihood estimation using robust standard errors. Marginal effects are presented. t-statistics in parentheses. The weights matrix has been row-normalized.

$* p<0.1, ** p<0.05, *** p<0.01$

**Table 9. Model Comparison of the estimated direct and spillover (marginal) effects on Life Satisfaction. Restricted sample to Toronto communities (N=274) (Weight Matrix Used: Inverse Distance).**

| | Dependent variable: Life Satisfaction | | | | |
|---|---|---|---|---|---|
| Independent (**Direct**) | OLS | SLX | SDEM | SDM | GNS |
| Household Income (log) | 0.137** | 0.154** | 0.156*** | 0.155*** | 0.151** |
| | (2.34) | (2.55) | (2.58) | (2.57) | (2.50) |
| Unemployment Rate | 0.001 | 0.001 | 0.001 | 0.001 | 0.001 |
| | (0.27) | (0.16) | (0.12) | (0.12) | (0.18) |
| Commute Duration | -0.007*** | -0.006*** | -0.006*** | -0.006*** | -0.006*** |
| | (-3.39) | (-2.95) | (-2.89) | (-2.85) | (-2.92) |
| Population Density (log) | -0.036*** | -0.038*** | -0.037*** | -0.038*** | -0.039*** |
| | (-5.16) | (-4.72) | (-4.65) | (-4.69) | (-4.93) |
| Proportion of Religious | 0.382*** | 0.342*** | 0.341*** | 0.340*** | 0.341*** |
| | (3.15) | (2.69) | (2.67) | (2.68) | (2.73) |
| Permanent Location (5y) | 0.140 | 0.109 | 0.103 | 0.100 | 0.105 |
| | (1.04) | (0.82) | (0.78) | (0.75) | (0.78) |
| Proportion of 4y degree | 0.115 | 0.101 | 0.097 | 0.098 | 0.104 |
| | (0.79) | (0.67) | (0.65) | (0.65) | (0.70) |
| Proportion Foreign Born | -0.224*** | -0.205** | -0.201** | -0.203** | -0.213** |
| | (-2.90) | (-2.24) | (-2.18) | (-2.23) | (-2.40) |
| Std Dev. of Life Satisfaction | -0.580*** | -0.577*** | -0.576*** | -0.577*** | -0.579*** |
| | (-11.97) | (-12.59) | (-12.58) | (-12.58) | (-12.62) |
| Independent (**Indirect**) | | | | | |
| Household Income (log) | | -0.082 | -0.085 | -0.116 | -0.163 |
| | | (-1.51) | (-1.55) | (-0.91) | (-1.00) |
| Unemployment Rate | | -0.004 | -0.003 | 0.000 | 0.007 |
| | | (-0.14) | (-0.12) | (0.01) | (0.19) |
| Std Dev. of Life Satisfaction | | 0.570 | 0.581 | 0.540 | 0.438 |
| | | (1.56) | (1.58) | (1.32) | (0.96) |
| P | | | | 0.072 | 0.174 |
| | | | | (0.31) | (0.68) |
| Λ | | | 0.081 | | -0.319 |
| | | | (0.22) | | (-0.66) |
| Adj. R-squared: | 0.686 | 0.689 | 0.688 | 0.689 | 0.689 |

Note: Maximum Likelihood estimation using robust standard errors. Marginal effects are presented. t-statistics in parentheses. The weights matrix has been normalized by dividing each element with the largest eigenvalue of the matrix.

\* $p<0.1$, \*\* $p<0.05$, \*\*\* $p<0.01$

**Table 10. Model Comparison of the estimated direct and spillover (marginal) effects on Life Satisfaction. Restricted sample to Ontario communities (N=355) (Weight Matrix Used: 1st Order B.C. Rook).**

| | Dependent variable: Life Satisfaction | | | | |
|---|---|---|---|---|---|
| Independent (**Direct**) | OLS | SLX | SDEM | SDM | GNS |
| Household Income (log) | 0.167*** | 0.183*** | 0.183*** | 0.197*** | 0.212*** |
| | (3.20) | (3.17) | (3.17) | (3.32) | (3.42) |
| Unemployment Rate | 0.005 | -0.000 | -0.000 | 0.000 | 0.001 |
| | (0.81) | (-0.01) | (-0.01) | (0.03) | (0.21) |
| Commute Duration | -0.007*** | -0.006*** | -0.006*** | -0.006*** | -0.005*** |
| | (-3.85) | (-3.83) | (-3.83) | (-3.32) | (-2.65) |
| Population Density (log) | -0.036*** | -0.037*** | -0.037*** | -0.035*** | -0.034*** |
| | (-5.26) | (-6.14) | (-6.14) | (-5.66) | (-5.14) |
| Proportion of Religious | 0.406*** | 0.407*** | 0.407*** | 0.392*** | 0.370*** |
| | (3.67) | (3.64) | (3.63) | (3.47) | (3.40) |
| Permanent Location (5y) | 0.205* | 0.127 | 0.126 | 0.112 | 0.113 |
| | (1.67) | (1.01) | (1.01) | (0.89) | (0.91) |
| Proportion of 4y degree | 0.114 | 0.074 | 0.074 | 0.068 | 0.072 |
| | (0.83) | (0.55) | (0.55) | (0.50) | (0.56) |
| Proportion Foreign Born | -0.239*** | -0.259*** | -0.259*** | -0.244*** | -0.223*** |
| | (-3.28) | (-3.35) | (-3.35) | (-3.09) | (-2.88) |
| Std Dev. of Life Satisfaction | -0.547*** | -0.546*** | -0.546*** | -0.545*** | -0.545*** |
| | (-11.94) | (-12.84) | (-12.84) | (-12.83) | (-12.82) |
| Independent (**Indirect**) | | | | | |
| Household Income (log) | | -0.011 | -0.011 | -0.044 | -0.083 |
| | | (-0.99) | (-0.98) | (-1.14) | (-1.22) |
| Unemployment Rate | | 0.012* | 0.012* | 0.010 | 0.008 |
| | | (1.70) | (1.71) | (1.51) | (1.09) |
| Std Dev. of Life Satisfaction | | 0.021 | 0.021 | 0.010 | 0.012 |
| | | (0.28) | (0.28) | (0.12) | (0.16) |
| $\rho$ | | | | -0.054 | 0.120 |
| | | | | (-1.13) | (1.13) |
| $\lambda$ | | | 0.002 | | -0.110 |
| | | | (0.02) | | (-0.80) |
| Adj. R-squared: | 0.655 | 0.659 | 0.659 | 0.660 | 0.660 |

Note: Maximum Likelihood estimation using robust standard errors. Marginal effects are presented. t-statistics in parentheses. The weights matrix has been row-normalized.

* p<0.1, ** p<0.05, *** p<0.01

**Table 11. Model Comparison of the estimated direct and spillover (marginal) effects on Life Satisfaction. Restricted sample to Ontario communities (N=355) (Weight Matrix Used: Inverse Distance).**

| | Dependent variable: Life Satisfaction | | | | |
|---|---|---|---|---|---|
| Independent (**Direct**) | OLS | SLX | SDEM | SDM | GNS |
| Household Income (log) | 0.167*** | 0.167*** | 0.154*** | 0.167*** | 0.151*** |
| | (3.20) | (2.93) | (2.75) | (2.92) | (2.72) |
| Unemployment Rate | 0.005 | 0.006 | 0.006 | 0.006 | 0.006 |
| | (0.81) | (1.01) | (1.16) | (1.03) | (1.12) |
| Commute Duration | -0.007*** | -0.006*** | -0.007*** | -0.006*** | -0.006*** |
| | (-3.85) | (-3.56) | (-3.77) | (-3.57) | (-3.66) |
| Population Density (log) | -0.036*** | -0.036*** | -0.037*** | -0.036*** | -0.037*** |
| | (-5.26) | (-5.53) | (-5.75) | (-5.42) | (-5.86) |
| Proportion of Religious | 0.406*** | 0.402*** | 0.408*** | 0.401*** | 0.410*** |
| | (3.67) | (3.44) | (3.62) | (3.43) | (3.71) |
| Permanent Location (5y) | 0.205* | 0.207* | 0.238* | 0.213* | 0.233* |
| | (1.67) | (1.66) | (1.94) | (1.71) | (1.87) |
| Proportion of 4y degree | 0.114 | 0.125 | 0.148 | 0.126 | 0.153 |
| | (0.83) | (0.93) | (1.13) | (0.94) | (1.19) |
| Proportion Foreign Born | -0.239*** | -0.231*** | -0.247*** | -0.231** | -0.253*** |
| | (-3.28) | (-2.57) | (-2.87) | (-2.56) | (-3.00) |
| Std Dev. of Life Satisfaction | -0.547*** | -0.547*** | -0.550*** | -0.546*** | -0.552*** |
| | (-11.94) | (-12.73) | (-12.80) | (-12.64) | (-12.83) |
| Independent (**Indirect**) | | | | | |
| Household Income (log) | | -0.014 | 0.001 | 0.008 | -0.041 |
| | | (-0.33) | (0.04) | (0.10) | (-0.42) |
| Unemployment Rate | | -0.010 | -0.015 | -0.013 | -0.008 |
| | | (-0.44) | (-0.74) | (-0.57) | (-0.31) |
| Std Dev. of Life Satisfaction | | 0.142 | 0.068 | 0.186 | -0.074 |
| | | (0.48) | (0.25) | (0.58) | (-0.20) |
| ρ | | | | -0.072 | 0.132 |
| | | | | (-0.30) | (0.60) |
| λ | | | -0.471 | | -0.695* |
| | | | (-1.23) | | (-1.76) |
| Adj. R-squared: | 0.655 | 0.655 | 0.655 | 0.655 | 0.655 |

Note: Maximum Likelihood estimation using robust standard errors. Marginal effects are presented. t-statistics in parentheses. The weights matrix has been normalized by dividing each element with the largest eigenvalue of the matrix.

* p<0.1, ** p<0.05, *** p<0.01

# Conclusion

Although there have long been strong theoretical arguments for the analysis of social comparisons of happiness and well-being between people and areas, there are only few relevant empirical studies. However, a recent publication in this journal [21] paves the way for inter-community comparisons to become now possible. In this article, we modified the standard utility models that are typically developed and used by economists, allowing us to explore the spatial spillover effects of life satisfaction between community areas in Canada. We empirically estimated the model using spatial econometric techniques. By applying these techniques, we are able to explore the extent to which characteristics of one community (e.g. unemployment rate) can affect the life satisfaction of a neighbouring community. Furthermore, the impact of a community's life satisfaction on the life satisfaction of neighbouring communities can also be assessed. Thus, we have used life satisfaction data at the community area level in Canada in order to explore possible spatial interdependencies. Overall, we examined possible spatial spillover effects in communities' happiness. Using data on life satisfaction, we proxy the utility of the communities and we empirically examine whether proximity plays any role.

Our results suggest that communities impact the life satisfaction levels of their neighbouring communities. The drivers of the interdependencies found are not limited to the characteristics of the neighbouring communities. Apart from the income level and unemployment rate, the level of satisfaction of a community affects the satisfaction level of its neighbours. These findings explain the clusters of high-high and low-low in well-being communities found in Canada. The results suggest that there is some evidence of spatial interdependencies between communities and that policy makers should take that into account when examining the well-being in rural and urban places. In particular, based on descriptive statistics, we observe that urban areas exhibit lower levels of life satisfaction, which is consistent with previous research [38,39]. Moreover, the dispersion of life satisfaction in those areas is higher as higher life satisfaction inequality is found in urban areas. Life satisfaction inequality is measured by the standard deviation of Life Satisfaction variable.

Furthermore, we found positive spatially autocorrelated clusters of regions with similar levels of live satisfaction being neighbours (e.g. positive-positive or negative-negative). The most persistent result from our spatial regressions is the significance of the indirect unemployment effect. In particular this finding suggests a possible 'could be worse' effect regarding unemployment: higher unemployment in neighbouring areas is associated with higher levels of community level subjective well-being. This highlights the possible important role of unemployment in shaping the happiness not only of individuals but also of communities and even of neighbouring communities. In addition, it illustrates the potential of our methodological approach to make a contribution to debates about socio-spatial comparisons and well-being between people and places.

Nevertheless, our analysis would have been more powerful if we had access to small area microdata with individual addresses that could have enabled us to perform a similar analysis of spatial interdependencies between actual neighbours. In addition to the paucity of such suitable microdata an additional methodological challenge of such an exercise would have been the computational intensity and computer memory requirements and management, given that a spatial matrix of the total population of Canada would be needed.

Moreover, we acknowledge the limitations of using a cross-sectional dataset instead of panel data. Using panel data we could examine the dynamics in the relationships we are interested in. Having only contemporaneous relations may be misleading but they still provide the interdependencies we would have expected between communities' life satisfaction. We should be cautious when we interpret such results and further research is needed with more disaggregated regional and panel data.

Overall, the work reported in this article presents a new framework that can be used to explore and quantify the extent to which community area happiness measured by individuals' measures of subjective life satisfaction may be affected by neighbouring communities. Similar patterns could be explored regarding other variables and considering issues pertaining to other economic variables (e.g. economic growth in neighbouring regions) but also (and especially given the current Covid-19 crisis) health-related variables (e.g. how the high prevalence of Covid-19 in an area may affect subjective happiness or sense of anxiety in neighbouring areas). The methodological framework specified in this article can be used and adapted to explore a wide range of regional and sub-regional socio-spatial interdependencies

## Acknowledgements:

The authors are grateful to John Helliwell, Hugh Shiplett and Christopher P. Barrington-Leigh for the public use dataset they created. An earlier version of this article was presented in a special session on spatial econometrics at the 59$^{th}$ European Regional Science Association Conference. We are very grateful to the participants of that session for their useful comments and suggestions. All responsibility for the analysis and interpretation of the data presented in this article lies with the authors

8. Krekel C, Kolbe J, Wuestemann H. The Greener, The Happier? The Effect of Urban Land Use on Residential Well-Being. Ecol Econ. 2016; 121: 117-27.

9. Frank R. Luxury Fever: Money and Happiness in an era of excess. New York: Princeton University Press; 1999.

10. Frank R. Falling Behind: How Rising Inequality Harms the Middle Class. Wildavsky Forum Series; 2007.

11. Dorling D. Injustice: Why social inequality persists. Bristol: Policy Press; 2011

12. Wilkinson R, Pickett K. The Spirit Level: Why Equality is Better for Everyone. London: Penguin; 2010.

13. Wilkinson R, Pickett K. The Inner Level: How More Equal Societies Reduce Stress, Restore Sanity and Improve Everyone's Well-being. London: Penguin; 2018.

14. Clark AE. Unemployment as a social norm: Psychological evidence from panel data. J Labor Economics. 2003; 21:323-351.

15. Powdthavee N. Are there geographical variations in the psychological cost of unemployment in South Africa? Soc Indic Res. 2007; 80(3):629-652.

16. Ballas D, Tranmer M. Happy People or Happy Places? A Multi-Level Modelling Approach to the Analysis of Happiness and Well-Being. Int Reg Sci Rev. 2012; 35:70-102.

17. Luttmer EFP. Neighbors as Negatives: Relative Earnings and Well-Being. Q J Econ. 2005; 120(3):963-1002.

18. Clark AE, Frijters P, Shields M. Relative Income, Happiness and Utility: An Explanation for the Easterlin Paradox and Other Puzzles. J Econ Literature. 2008; 46:95–144.

19. Aslam A, Corrado L. The geography of well-being. J Econ Geogr. 2012; 12:627-649.
45